\documentstyle[twoside,12pt,epsf]{article}
\let\ssection=\section
\renewcommand{\section}{\setcounter{equation}{0}\ssection}

\def\head#1#2{
\markboth{}{}
\setcounter{page}{#1}
\setcounter{section}{0}
\begin{huge}
\begin{flushleft}
\noindent\hangindent\parindent
{#2}
\end{flushleft}

\end{huge}
\bigskip
}

\def\auaddr#1#2{
{\noindent\LARGE\em#1}
\medskip

{\noindent #2}
\medskip
}

\def\and{
{\LARGE\em \&}
\bigskip
}

\def \half {\raise1pt\hbox{$\scriptstyle
        {1 \over 2}\displaystyle$}}
\def \twentyfour {\raise1pt\hbox{$\scriptstyle
       {1 \over {24}}\displaystyle$}}
\def \sixth {\raise1pt\hbox{$\scriptstyle
       {1 \over 6}\displaystyle$}}

   \let\qd=\quad
\let\a=\alpha \let\be=\beta \let\g=\gamma \let\de=\delta
   
   \let\m=\mu
\let\n=\nu  \let\r=\rho 
\let\om=\omega 
\let\ph=\varphi  \let\PH=\Phi 
  
 \let\Ga=\Gamma 
\def\0#1#2{\frac{#1}{#2}}
\def\s0#1#2{\mbox{\small{$\frac{#1}{#2}$}}}
\def\5{\bar }  \def\6{\partial } \def\7{\hat } \def\4{\tilde }
 
\def\bea{\begin{eqnarray}} \def\eea{\end{eqnarray}}
\def\beann{\begin{eqnarray*}} \def\eeann{\end{eqnarray*}}
\def\beq{\begin{equation}} \def\eeq{\end{equation}}
\def\ba{\begin{array}} \def\ea{\end{array}}
  
\def\cA{{\cal A}}  
 \def\cM{{\cal M}} 
\def\cF{{\cal F}}  \def\cH{{\cal H}}
  
\def\cX{{\cal X}}

 \def\f#1#2#3{{f_{#1#2}}^{#3}}

\newcommand{\mysection}[1]{\section{#1}\setcounter{equation}{0}}
\def\Gl#1{(\ref{#1})}
\def\ben{\begin{enumerate}}
\def\een{\end{enumerate}}
\def\lb{\left(} \def\rb{\right)}
\begin{document}
{\thispagestyle{empty}
\hspace*{\fill} NIKHEF--H 94--16\\
\hspace*{\fill} KUL--TF--94/17\\
\hspace*{\fill} hep-th/9407061
\vspace*{2cm}

\begin{centering}
{\LARGE {\bf BRST--antibracket cohomology in\\[.1cm]
 2d conformal gravity\\[.1cm]}}\vspace{2cm}

{\renewcommand{\thefootnote}{\fnsymbol{footnote}}
\Large Friedemann Brandt\footnote{Supported by Deutsche
Forschungsgemeinschaft. E-mail:~t28@nikhef.nl}}\vspace{.5cm}

NIKHEF--H,~Postbus 41882,~NL--1009 DB Amsterdam,~The Netherlands\vspace{1cm}

{\renewcommand{\thefootnote}{\fnsymbol{footnote}}
{\Large
Walter Troost\footnote{Onderzoeksleider, NFWO, Belgium.
E--mail:~Walter\%tf\%fys@cc3.kuleuven.ac.be}}},
{\renewcommand{\thefootnote}{\fnsymbol{footnote}}
{\Large
Antoine Van Proeyen\footnote{Onderzoeksleider, NFWO, Belgium.
E--mail:~fgbda19@cc1.kuleuven.ac.be}}}\vspace{.5cm}

Instituut voor Theoretische Fysica,~Katholieke Universiteit Leuven,\\
Celestijnenlaan 200D,~B--3001 Leuven, Belgium\vspace{2cm}

To appear in the proceedings of the\\ {\it Geometry of Constrained
Dynamical Systems} workshop,\\ Isaac Newton Institute for
Mathematical Sciences, Cambridge, June 15--18, 1994

\end{centering}
}
\newpage

\head {1}{BRST-antibracket cohomology in 2d conformal gravity}

\auaddr{\renewcommand{\thefootnote}{\fnsymbol{footnote}}
Friedemann Brandt\footnote{Supported by Deutsche
Forschungsgemeinschaft. E-mail:~t28@nikhef.nl\\
Address after October 31, 1994:
Instituut voor Theoretische Fysica,~KU
Leuven,~Celestijnenlaan 200D,~B--3001 Leuven,~Belgium.}}
{NIKHEF--H,~Postbus 41882,~NL--1009 DB Amsterdam,~The Netherlands}

\auaddr{\renewcommand{\thefootnote}{\fnsymbol{footnote}}
Walter Troost\footnote{Onderzoeksleider, NFWO, Belgium.
E--mail:~Walter\%tf\%fys@cc3.kuleuven.ac.be},
\renewcommand{\thefootnote}{\fnsymbol{footnote}}
Antoine Van Proeyen\footnote{Onderzoeksleider, NFWO, Belgium.
E--mail:~fgbda19@cc1.kuleuven.ac.be}}
{Instituut voor Theoretische Fysica,~Katholieke Universiteit Leuven,\\
Celestijnenlaan 200D,~B--3001 Leuven, Belgium}

\begin{abstract}
We present results of a computation of the BRST-antibracket
cohomology in the space of local functionals of the fields and
antifields for a class of $2d$ gravitational theories which are
conformally invariant at the classical level. In particular
all classical local action functionals, all candidate
anomalies and all
BRST--invariant functionals depending nontrivially
on antifields are given and discussed for these models.
\end{abstract}
\setcounter{footnote}{0}
\mysection{Introduction}\label{intro}

Conformal invariance plays a crucial role in various
two dimensional physical models. Of special interest
is the question whether conformal invariance of a classical
theory is maintained in the quantum theory or becomes
anomalous. In string theory, for instance, vanishing
of the conformal anomaly determines the critical
dimension and imposes ``equations of motion'' in target
space \cite{friedan}. Since the work of Wess and Zumino
\cite{wz} it is well-known that anomalies have to satisfy
consistency conditions following from the algebra of the
symmetries of the classical theory. The general form of these
conditions can be elegantly
formulated in the BV-antifield formalism \cite{bv} as the
vanishing of the antibracket of the proper solution $S$
of the classical master equation and
a functional $\cA$ of the fields and antifields
representing the anomaly \cite{anbv}:
\beq (S,\cA)=0.\label{i1}\eeq
\Gl{i1} amounts to a cohomological problem since it requires
BRST invariance of $\cA$: the BRST operator\footnote{We call
$s$ the BRST operator although this terminology often is
used only when it acts on the fields, and not on
antifields.} $s$
is defined on arbitrary functionals $\cF$ of fields and antifields
through
\beq s\, \cF:=(S,\cF)\label{i2}\eeq
and its nilpotency is implied by the Jacobi identity for the
antibracket and by the fact that $S$ solves the classical master
equation
$(S,S)=0$:
\beq s^2=0. \label{i3}\eeq
Let us denote by $H^*(s)$ the
BRST cohomology in the relevant space of functionals of the fields and
antifields which must be specified
in each particular case (usually it is the space of local functionals
whose precise definition must be adapted to the problem).
Since the BRST operator increases the ghost
number ($gh$) by one unit due to $gh(S)=0$, $H^*(s)$ can be computed
in each subspace of functionals with a definite
ghost number $g$ separately where we denote it by $H^g(s)$.
Anomalies are represented by cohomology classes of $H^1(s)$, at least
if the ghost number is conserved in the quantum theory which holds
at tree level due to $gh(S)=0$. However it is often useful to
compute $H^g(s)$ for other values of $g$ as well. In particular
$H^0(s)$ is interesting since it contains $S$ itself. This opens
the possibility to construct $S$ by computing $H^0(s)$
after fixing the desired field content and gauge invariances
of a model. This was actually our starting point for
the computation of $H^*(s)$ in a class of two dimensional
models which are conformally invariant at the classical level
(see next section). A computation of $H^0(s)$ can also be
useful for given $S$ since it
can provide information about observables or counterterms arising in
a theory. Moreover antifield--dependent solutions $\cM$ of
$(S,\cM)=0$ with ghost number 0
can have interesting interpretations and applications. For instance
they are needed for the construction of
a functional $\4S=S+\cM+\ldots$ satisfying $(\4S,\4S)=0$. If $\4S$
exists, it provides a nontrivial extension of the theory
which is consistent
in the sense that it is invariant under suitable extensions of the
gauge transformations of the original theory characterized
by $S$. This was pointed out and exemplified in \cite{hen}.
BRST--invariant functionals $\cM$ with ghost number 0 which do not
satisfy $(\cM,\cM)=0$ may alternatively receive an interpretation
as background charges,
which can cancel anomalies. This can be implemented in the BV
antifield formalism \cite{hiding} by
formally considering $\cM=\sqrt{\hbar}\cM_{1/2}$ as a
contribution to the quantum
action $W$ of order $\sqrt{\hbar}$ since
$W=S+\sqrt{\hbar}\cM_{1/2}+\hbar\cM_1+\ldots$ implies
$(W,W)=\hbar(\cM_{1/2},\cM_{1/2})+\ldots$, i.e. $(\cM,\cM)$ indeed
can cancel one-loop anomalies.

\mysection{Characterization of the models}\label{char}
Our aim was the computation of $H^*(s)$
for a class of two dimensional models
which are conformally invariant at the classical level.
To this end we did not characterize these models
by specific conformally
invariant classical actions but we only specified
the field content and
the gauge invariances of the classical theory.
Then we computed $H^*(s)$ in the space
of antifield--independent
functionals which for ghost number 0 in particular
provides the most general local classical action functional $S_0$ and
thus characterizes more precisely
the models to which our results apply. Then we
completed the computation of $H^*(s)$ by inclusion of the antifields.
This procedure is possible due to the closure of the algebra of
gauge symmetries since in this case the BRST transformations of the
fields do not depend on the antifields. The BRST
transformations of the antifields however involve (functional
derivatives of) $S_0$ and therefore their inclusion requires
the knowledge of $S_0$.

In detail, the models which we investigated are characterized by
\ben
\item[(i)] Field content: $S_0$ is a
local functional of the $2d$ metric $g_{\a\be}=g_{\be\a}$
($\a,\be\in\{+,-\}$)
and a set of bosonic scalar matter fields $X^\m$
($\m\in\{1,2,\ldots,D\}$).
\item[(ii)] $S_0$ is invariant under $2d$ diffeomorphisms and
local Weyl transformations of the metric $g_{\a\be}$.
\item[(iii)] $S_0$ does not possess any
nontrivial gauge symmetries apart from those mentioned in
(ii).\footnote{Trivial gauge symmetries
of an action $S_0$ depending on a set of (bosonic) fields
$\ph^i$ are by definition
of the form $\de_\epsilon\ph^i=P^{ij}\de S_0/\de\ph^j$
where $P^{ij}=-P^{ji}$ are arbitrary functions of the $\ph^i$,
arbitrary parameters $\epsilon(x)$
and the derivatives of the $\ph^i$ and $\epsilon$.}
\een
In order to make (i) precise we have to add the definition of
local functionals we used:
\ben
\item[(iv)] A functional of a set
of fields $Z^A$ is called local if its integrand is
a polynomial in the derivatives of the $Z^A$
(without restriction on the order of derivatives)
but may depend nonpolynomially
on the undiffentiated fields $Z^A$ and explicitly on the coordinates
$x^\a$ of the two dimensional base manifold.
\een
(ii) requires
\beq s\, S_0=0 \label{i4}\eeq
where $s$ acts on $g_{\a\be}$ and $X^\m$ according to
\bea s\, g_{\a\be}=\xi^\g\6_\g g_{\a\be}+ g_{\g\be}\6_\a\xi^\g
+g_{\a\g}\6_\be\xi^\g+c\, g_{\a\be} ,\qd
s\, X^\m=\xi^\a\6_\a X^\m. \label{i5}\eea
Of course these are just the BRST transformations of $g_{\a\be}$ and
$X^\m$ where $\xi^\a$ and $c$ are the anticommuting ghosts of
diffeomorphisms
and local Weyl transformations respectively. The
BRST transformations of the ghosts are chosen such that
\Gl{i3} holds on all fields $g_{\a\be},X^\m,\xi^\a,c$. This leads to
\beq s\, \xi^\a=\xi^\be\6_\be\xi^\a,\qd s\, c=\xi^\a\6_\a
c\, .\label{i6}\eeq
\Gl{i4}--\Gl{i6} and requirement (iii) guarantee
that the proper solution of the classical master equation is given by
\beq S=S_0-\int d^2x\, (s\PH^A)\PH^*_A \label{i7}\eeq
where we used customary collective notations
$\{\PH^A\}=\{g_{\a\be},X^\m,\xi^\a,c\}$ and
$\{\PH^*_A\}=\{g^{*\a\be},X^*_\m,\xi^*_\a,c^*\}$ for fields and
antifields.
The BRST transformation of $\PH^*_A$ is given by the functional right
derivative of $S$ w.r.t.~$\PH^A$ (in our conventions the BRST operator
acts from the left everywhere)
\beq s\, \PH^*_A=\0 {\de_r S}{\de\PH^A}\, .\label{i8}\eeq
{\it Remark:}\\
Although we allow the integrands of local functionals to
depend explicitly on $x^\a$ according to (iv), it turns out that
integrands of BRST--invariant functionals actually do
not carry an explicit $x$-dependence (up to trivial
contributions of course). Nevertheless we need
this definition of local functionals in order to cancel candidate
anomalies
as e.g.~$\int d^2x\, \xi^\a L$ where $L$ is a Weyl invariant density
($sL=
\6_\a(\xi^\a L)$). Namely these functionals are BRST invariant but
not BRST exact unless we admit counterterms whose integrands
depend explicitly (and in fact polynomially) on the $x^\a$.
These are well-known features of all gravitational theories
(cf. \cite{bon,grav,ten}). However, if one takes into account
topological
properties of the base manifold, the $x$-independence
of the integrands of nontrivial BRST--invariant functionals holds
strictly only if the manifold does not allow closed
$p$-forms with $p\neq 0$ which are not exact. The results we
present in the next section therefore hold in a strict sense only
under this additional
assumption (cf.~\cite{ten} for general remarks on this point).

\mysection{Results}\label{results}
\subsection{Antifield--independent functionals}
We found that $H^g(s)$ vanishes for $g>4$ in the space of local
antifield--independent functionals, i.e.~each
local BRST--invariant functional with ghost number $g>4$
which does not depend on antifields is the BRST variation of
a local functional with ghost number $g-1$ which also does not
depend on antifields. We only spell out the results for
$g=0,1$. Those for $g=2,3,4$ will be given in \cite{paper}.

$H^0(s)$ provides the most general classical action $S_0$. It is given
by
\beq S_0=\int d^2x\, \left( \s0 12\sqrt{g}\,
g^{\alpha\beta}G_{\mu\nu}(X)
\partial_\alpha X^\mu
\partial_\beta X^\nu
+B_{\mu\nu}(X)\partial_+ X^{\mu}\partial_- X^{\nu}\right)
\label{r1}\eeq
with $g=|\det(g_{\a\be})|$. $G_{\m\n}$ and $B_{\m\n}$
are arbitrary functions of the $X^\m$ satisfying
\[ G_{\m\n}=G_{\n\m},\qd B_{\m\n}=-B_{\n\m}.\]
$B_{\m\n}$ is defined only up to contributions $\6_\m B_\n(X)
-\6_\n B_\m(X)$ which yield
total derivatives in the integrand of \Gl{r1}. Here and henceforth
\[ \6_\m:=\0 \6{\6X^\m} \]
denote derivatives w.r.t. matter fields.
\Gl{r1} is the most general functional
satisfying requirements (i) and (ii) listed in section \ref{char}
with the restrictions imposed by (iv).
(iii) represents an additional requirement which excludes
e.g.~functions
$G_{\m\n}$ and $B_{\m\n}$ admitting a nonvanishing solution
$g^\m(X)$ of
\beq G_{\m\n}g^\n=\Ga_{\m\n\r}g^\r=H_{\m\n\r}g^\r=0\label{r2}\eeq
where
\beq \Ga_{\m\n\r}=
\s0 12\left(\6_\m G_{\n\r}+\6_\n G_{\m\r}-\6_\r G_{\m\n}\right),\qd
H_{\m\n\r}=\6_\m B_{\n\r}+\6_\n B_{\r\m}+\6_\r B_{\m\n}.\label{r3}
\eeq
Namely \Gl{r2} implies the invariance of
$S_0$ under $\de_\epsilon X^\m=\epsilon(x) g^\m(X)$ for
an arbitrary function $\epsilon(x)$ and thus the presence of an
additional
gauge invariance which violates requirement (iii).

It is also worth noting that invariance of the theory under target
space reparametrizations can be
elegantly formulated in the antifield formalism as well
(this kind of an invariance must not be confused
with invariance of the action functional in the
usual sense, of course).
Namely any two action functionals
$S'_0[X]:=S_0[X+\de X]$ and $S_0[X]$ which are
related by an arbitrary infinitesimal target space
reparametrization $\de X^\m=f^\m (X)$ differ by the
BRST-variation of a local antifield--dependent functional
(with BRST transformation
of the antifields defined by means of $S_0[X]$).
Both $S'_0[X]$ and $S_0[X]$ are of the form \Gl{r1} with
functions $G'_{\m\n}$ and $G_{\m\n}$ resp. $B'_{\m\n}$ and $B_{\m\n}$
related by\footnote{(Anti-)Symmetrization of indices
is defined by $f_{(\m\n)}=\s0 12(f_{\m\n}+f_{\n\m})$ etc.}
\begin{equation}
G'_{\mu\nu}=G_{\mu\nu}+2\partial_{(\mu}f_{\nu)}-\Gamma_{\mu\nu\rho}
f^\rho\
;\qquad B'_{\mu\nu}=B_{\mu\nu}+2\partial_{[\mu}B_{\nu]}
+H_{\mu\nu\rho}f^\rho     \label{transfGB}
\end{equation}
where $f_\m:=G_{\m\n}f^\n$. We note that an analogous statement
holds for {\it any} theory
characterized by a local action functional
$S_0[\phi]$ (where $\phi^i$ are the fields with
ghost number 0).
Namely consider infinitesimal (local) field redefinitions
$\de\phi^i=f^i(\phi,\6\phi,\ldots)$ which are chosen such that
\beq s\, \lb S_0[\phi+\de\phi]-S_0[\phi]\rb=0\label{2}\eeq
holds with the BRST operator encoding the gauge symmetries
of $S_0[\phi]$. Then there is a (local)
functional $\Ga^{-1}$ with ghost number $-1$ such that
\beq S_0[\phi+\de \phi]-S_0[\phi]=s\, \Ga^{-1}[\PH,\PH^*].
\label{1}\eeq
This statement holds also for theories without gauge
invariances.\footnote{Notice that transformations
$\de\phi^i$ satisfying \Gl{2} are more general transformations than
symmetry transformations of $S_0$
since the latter satisfy the much stronger condition
$S_0[\phi+\de\phi]=S_0[\phi]$.}
In this case $s$ reduces to the Koszul-Tate differential
$\de_{KT}=\int d^2x(\de S_0[\phi]/\de\phi^i)\de/\de\phi^*_i$
and \Gl{1} holds obviously for arbitrary field
redefinitions $\de\phi^i$.

$H^1(s)$ provides the antifield--independent candidate anomalies. They
are given by
\bea \cA&=&\cH_+ +\cH_- + \cX_+ + \cX_-,\label{r4}\\
     \cH_\pm&=&a_\pm\int d^2x\, c^\pm(\6_\pm)^3h_{\mp\mp},\label{r5}\\
     \cX_\pm&=&\int d^2x \, \s0 1{1-y}
               (\6_\pm\xi^\pm+h_{\mp\mp}\6_\pm\xi^\mp)
\nabla_+X^\m\nabla_-X^\n f_{\m\n}^\pm(X)\label{r6}\eea
where $a_+,a_-$ are constants, $f^+_{\m\n},f^-_{\m\n}$ are arbitrary
functions of the $X^\m$ and
\beq\ba{ll}  h_{\pm\pm}=g_{\pm\pm}/(g_{+-}+\sqrt{g}),&
y=h_{++}h_{--},\\
\nabla_\pm X^\m=(\6_\pm-h_{\pm\pm}\6_\mp)X^\mu,
& c^\pm=\xi^\pm+h_{\mp\mp}\xi^\mp.\ea\label{r7}\eeq
Using the original components of the metric, $\cX_\pm$ read
\beann\cX_\pm=\int d^2x\, (\6_\pm\xi^\pm+h_{\mp\mp}\6_\pm\xi^\mp)
\left( \s0 12\sqrt{g}\, g^{\alpha\beta}f^\pm_{(\mu\nu)}(X)
\partial_\alpha X^\mu
\partial_\beta X^\nu\right.\\
\left.+f^\pm_{[\mu\nu]}(X)\partial_+ X^{\mu}\partial_- X^{\nu}\right).
\eeann
In fact the parts of $\cX_\pm$ containing
the symmetric and antisymmetric parts of $f^\pm_{\mu\nu}$ are
separately BRST invariant. They are also nontrivial and inequivalent
in the space of antifield--independent
functionals. We remark however that those functionals
$\cX_+, \cX_-$ which arise from contributions
\beq 2\6_{(\m} H^\pm{}_{\n)}\mp   2
\6_{[\m} H^\pm{}_{\n]} + (H_{\m \n \r}-2 \Ga_{\m \n \r})\, H^{\pm\r}
\qd (\mbox{with}\  H^\pm{}_{\m}:=G_{\m\n}H^{\pm\n})
\label{r8}\eeq
to $f^\pm_{\m\n}$ are trivial in the space of local functionals
of the fields and antifields where $H^{\pm\m}(X)$ are arbitrary
functions of the $X^\m$ and the upper (lower) sign refers to
contributions to
$f^+_{\mu\nu}$ ($f^-_{\mu\nu}$).

\subsection{Antifield--dependent functionals}

The existence and explicit form of antifield--dependent
BRST--invariant
functionals depends
of course on the specific form of $S_0$, i.e.~on the specific choice
of
the functions $G_{\m\n}$ and $B_{\m\n}$ in \Gl{r1} since they enter in
the
BRST transformation of $X^*_\m$ and $g^{*\m\n}$, see \Gl{i8}.
Nevertheless
one
can classify all antifield--dependent BRST--invariant
functionals as follows:
\ben
\item[a)] $g\not\in\{-1,0,1\}$:\\
There are no antifield--dependent cohomology classes
in these cases, independently of the specific
form of $S_0$. More precisely:
If $W^g$ is an antifield--dependent local BRST--invariant functional
with
ghost
number $g\not\in\{-1,0,1\}$ then there is a local functional
$W^{g-1}$ such that $\4W^g:=W^g-sW^{g-1}$ does not depend on
antifields
anymore.
\item[b)] $g=-1$:\\
BRST--invariant local
functionals with ghost number $-1$
exist if and only if $G_{\m\n}$ and $B_{\m\n}$ admit
a nonvanishing solution $f^\m(X)$ of
\beq \6_\m f_\n+\6_\n f_\m-2\Ga_{\m\n\r}f^\r=0,\qd
     H_{\m\n\r}f^\r=\6_\m H_\n -\6_\n H_\m
\label{r9}\eeq
for some arbitrary functions $H_\m(X)$.
\Gl{r9} identifies $f_\m:=G_{\m\n} f^\n$ as the components
of a Killing vector in target space. Any solution
of \Gl{r9} generates a continuous {\it global} symmetry of $S_0$
through
\beq \de_\epsilon X^\m=\epsilon\, f^\m(X),\qd
\epsilon=const.\label{r10}\eeq
In other words: BRST--invariant local
functionals with ghost number $-1$ correspond one-to-one
to these global symmetries of $S_0$. They are given by
\beq W^{-1}=\int d^2x\, X^*_\m f^\m(X).\label{r11}\eeq
\item[c)] $g=0$:  \\
BRST--invariant local
functionals with ghost number $0$ depending nontrivially on the
antifields exist if and only if $G_{\m\n}$ and $B_{\m\n}$ admit
a nonvanishing solution $f^\m(X)$ of
\bea 0=\6_\m f_\n+\6_\n f_\m-2\Ga_{\m\n\r}f^\r,\qd
     0=\6_\m f_\n -\6_\n f_\m\mp H_{\m\n\r}f^\r
\label{r12}\eea
where the second condition must be satisfied either with the $+$ or
the $-$ sign for a particular solution $f^\m$.
Comparing \Gl{r12} and \Gl{r9} we conclude that
\Gl{r12} requires that $S_0$ possesses a global symmetry \Gl{r10} with
the additional restriction imposed by the second condition \Gl{r12}.
The BRST--invariant functionals with ghost number 0 arising from a
solution of \Gl{r12} with a minus sign in front of $H_{\m\n\r}f^\r$
are given by
\beq \cM_+=\int d^2x\left[X_\mu ^*(\partial _+\xi ^+ + h_{--}\partial
_+\xi ^-) - \s0 2{1 - y}
\nabla_+X^\nu  \partial _+h_{--}  G_{\mu
\nu }\right]f^\mu \label{r12a}\eeq
and the functionals $\cM_-$ arising from a solution of \Gl{r12} with
a plus sign
in front of $H_{\m\n\r}f^\r$ are obtained from \Gl{r12a} by exchanging
all $+$ and $-$ indices. These solutions do not satisfy
$({\cal M},{\cal M})=0$ and can thus not be added to the extended
action without breaking $(S,S)=0$, however they can be used to
introduce background charges as explained in the introduction.
Namely, taking $h_{++} = 0$, dropping the corresponding $\xi ^-$
ghost, and
specialising to $G_{\mu \nu}  = \delta _{\mu \nu }$, \Gl{r12a} becomes
$\int d^2x \, (X_\mu ^* \partial _+ \xi ^+ - 2\partial _+ X_\mu
 \partial _+ h_{--})\, f^{\mu }$
in which one recognises the so-called background charge terms (see
\cite{hiding} for their inclusion in the BV formalism).
Therefore, \Gl{r12a} constitutes the generalization of this chiral
gauge treatment.
\item[d)] $g=1$:\\
In this case we obtain \Gl{r2} as necessary and sufficient
conditions for the existence of BRST--invariant local
functionals with ghost number $1$ depending nontrivially on the
antifields. As discussed above, \Gl{r2} implies that $S_0$
possesses an additional
gauge symmetry which violates requirement (iii) and thus has to be
excluded. Namely in presence of additional gauge symmetries,
\Gl{i7} is not a proper solution of the classical master equation
anymore. To construct a proper solution one must introduce a ghost
and its antifield for each additional gauge symmetry. In the
extended space
of functionals depending also on these additional fields,
the antifield--dependent functionals arising from
solutions $g^\m$ of \Gl{r2} indeed are trivial. Nevertheless
one of course has to reexamine the whole investigation of $H^*(s)$
in the case of a higher gauge symmetry
and therefore our results do not apply to this case.
\een

\mysection{Sketch of the computation}\label{meth}

In the first step of the computation, the BRST cohomology in the space
of local functionals $W=\int d^2x f$ is related to the
BRST cohomology in the space of local functions by means of the
descent equations following from $sW=0$:
\beq s\,\om_2+d\,\om_1=0,\qd s\,\om_1+d\,\om_0=0,\qd s\,\om_0=0
\label{m1}
\eeq
where $\om_2=d^2x f$ is the integrand of $W$
written as a 2-form and $\om_1$ and $\om_0$ are local 1- and 0-forms.
It is well-known that the descent equations terminate
in gravitational theories always with a nontrivial 0-form $\om_0$
(contrary to the Yang--Mills case)\footnote{This statement
holds in a strict sense only in absence of closed $p$-forms ($p\neq
0$)
which are not exact,~cf.~\cite{ten}.} and that their ``integration''
is trivial:
\beq \om_1=b\,\om_0,\qd \om_2=\s0 12\, bb\,\om_0,\qd b=
dx^\a \0 \6{\6\xi^\a}.\label{m2}\eeq
According to these statements which
were first proved and applied in
\cite{grav} (for arbitrary dimensions) it is sufficient
to determine the general solution of
\beq s\,\om_0=0\label{m3}\eeq
in the space of local functions of the fields and their derivatives.
The BRST--invariant functionals resp.~their integrands are then
obtained via \Gl{m2} from the solutions of \Gl{m3}.

The investigation of \Gl{m3} is considerably simplified by
performing it in an appropriate new basis of variables
substituting the fields, antifields and their
derivatives. The construction of this new basis is
the second and crucial step within the computation.
The new basis contains in particular the following
variables $T_{m,n}^\m$
substituting one-by-one the partial derivatives
$(\6_+)^m(\6_-)^nX^\m$ of the matter fields:
\beq T_{m,n}^\m=\lb \0 \6{\6c^+}\, s\rb^m\lb \0 \6{\6c^-}\, s\rb^n
X^\m
\label{m4}\eeq
where $c^+$ and $c^-$ are the ghost variables defined in
\Gl{r7} and it is understood that the BRST transformations
occurring in \Gl{m4}
are expressed in terms of these ghosts. The first few
(and most important) $T$'s are given by
\[ T_{0,0}^\m=X^\m,\qd T_{1,0}^\m=\s0 1{1-y}\nabla_+X^\m,\qd
T_{0,1}^\m=\s0 1{1-y}\nabla_-X^\m\]
with $\nabla_\pm X^\m$ as in \Gl{r7}.
The most important ghost variables are
\beq c^n_+=\s0 1{(n+1)!}\, (\6_+)^{n+1}\, c^+,\qd
c^n_-=\s0 1{(n+1)!}\, (\6_-)^{n+1}\, c^- ,\qd n\geq -1.\label{m6}\eeq
The remarkable
property of the $T_{m,n}^\m$ is that they span the
representation space for two copies of the
``Virasoro algebra'' (without central extension) whose
associated ghosts are just the variables \Gl{m6}. Namely
the BRST transformations of $T_{m,n}^\m$ and $c^n_\pm$ can
be written as
\beq s\, T_{m,n}^\m=\sum_{k\geq -1} (
c^k_+L_k^++c^k_-L_k^-)T_{m,n}^\m,\qd
s\, c^k_\pm=\s0 12\f mnk c^m_\pm c^n_\pm\label{m9}\eeq
where $L^+_n$ and $L^-_n$ represent on the $T_{m,n}^\m$ the
Virasoro algebra according to
\beq {}[L_m^\pm,L_n^\pm]=\f mnk L_k^\pm,
\qd [L_m^+,L_n^-]= 0,\qd \f mnk=(m-n)\de_{m+n}^k\qd (m,n,k\geq -1).
\label{m7}\eeq
$L^\pm_k T_{m,n}^\m$ can be evaluated using \Gl{m7} and
\beq T_{m,n}^\m=\lb L_{-1}^+\rb^m\lb L_{-1}^-\rb^n X^\m,\qd
L_n^\pm X^\m=0\qd \forall n\geq 0.\label{m7a}\eeq
The equivalence of \Gl{m4} and the first relation \Gl{m7a}
can be verified using $s^2=0$ and the following representation
of $L_n^\pm$ on $T_{m,n}^\m$ which is implied by \Gl{m9}:
\beq L^+_n=\left\{s,\0 \6{\6c^n_+}\right\},\qd
L^-_n=\left\{s,\0 \6{\6c^n_-}\right\},\qd n\geq -1.\label{m5}\eeq
The $T_{m,n}^\m$ are called tensor fields.
In fact one can extend the definition of
tensor fields to the antifields. Of particular importance are
those tensor fields which substitute $X^*_\m$. They are given by
\beq \7X^*_\m=\s0 1{1-y}\, X^*_\m.\label{m10}\eeq

In the third step
one proves by means of standard methods
that nontrivial contributions to solutions of \Gl{m3}, written
in terms of the new basis,
depend on the fields, antifields and their derivatives
only via the $c^n_\pm$ and the tensor fields
constructed of the matter fields and the
antifields since all other variables group into trivial
systems of the form $(a,sa)$ and do not enter in \Gl{m9}.

In step four we take advantage of the fact that
$L_0^+$, $L_0^-$ are diagonal on all tensor fields
and on the ghosts \Gl{m6} on which these generators
are defined by means of \Gl{m5}. Namely one has
\beq L_0^+T_{m,n}^\m=mT_{m,n}^\m,\
  L_0^-T_{m,n}^\m=nT_{m,n}^\m,\
 L_0^\pm c^n_\pm=nc^n_\pm,\  L_0^\pm c^n_\mp=0
\label{m8}\eeq
and similar relations for the tensor fields constructed of the
antifields (e.g.~$\7X^*_\m$ has $(L_0^+,L_0^-)$ weights $(1,1)$).
By means of standard arguments one concludes
that solutions of \Gl{m3} can be assumed to have total weight $(0,0)$
(all other contributions to $\om_0$ are trivial).

The fifth and final step consists in the investigation of
\Gl{m3} in the space of those local functions of the ghosts
\Gl{m6} and the tensor fields which have total weight
zero under both $L_0^+$ and $L_0^-$. It turns out that
$c^{-1}_+=c^+$ and $c^{-1}_-=c^-$ are the only variables
having negative weights under $L_0^+$ or $L_0^-$. In
fact they have weights $(-1,0)$ and $(0,-1)$ respectively.
Since the ghosts anticommute, there are only
few possibilities to construct local functions
with total weight $(0,0)$ at all. In fact the whole
computation reduces to the investigation of functions
of the following quantities:
\bea & &
c_\pm^0 ,\qd
\tilde c_\pm\equiv 2 c_\pm^{-1}c_\pm^1,\qd
X^\mu=T^\mu_{0,0},\qd
T^\mu_ + \equiv c_+^{-1} T^\mu_{1,0},\qd
T^\mu_ - \equiv c_-^{-1} T^\mu_{0,1},\nonumber\\
& &
T_{+-}^\mu\equiv c_-^{-1}c_+^{-1} T^\mu_{1,1},\qd
T^*_\mu\equiv c_+^{-1}c_-^{-1} \7X^*_\mu
\label{m12}\eea
on which $s$ acts according to
\bea & & s\, c_\pm^0  =  \tilde c_\pm,\
 s \, X^\mu=  T^\mu_+ +T^\mu_-,\
s\, T^\mu_ +  = T_{+-}^\mu,\
s\, T^\mu_ -  = -T_{+-}^\mu,\nonumber\\
& & s\, T^*_\mu = 2G_{\m\n}T^\n_{+-}+(H_{\n\r\m}-2\Ga_{\n\r\m})
T^\n_+T^\r_-\label{m13}\eea
where $sT^*_\mu $ follows from \Gl{i8} and thus of course requires
the knowledge of \Gl{r1} which is obtained from the solution of the
antifield independent problem. Taking into account the algebraic
identities relating the quantities \Gl{m12} as a
consequence of the odd grading of $c^\pm$, like
$\4c_+\4c_+ $=$T_+^\m\4c_+$=$T_+^\m T_+^\n$=$0$ etc., one
sees that the space of nonvanishing functions
of these quantities is rather small apart
from the occurrence of arbitrary functions of
$T_{0,0}^\m=X^\m$.
This allows ultimately to solve \Gl{m3} completely.
The solution of \Gl{m3} which yields \Gl{r1} is
for instance given by $T_+^\m T_-^\n K_{\m\n}(X)$
where the symmetric and antisymmetric parts of
$K_{\m\n}$ are just $G_{\m\n}$ resp.~$B_{\m\n}$.

\mysection{Summary}
We have determined the complete BRST-antibracket cohomology
in the space of local functionals for
theories satisfying the assumptions (i)--(iv) listed in
section \ref{char}. We found that nontrivial cohomology classes exist
only
for ghost numbers $g=-1,\ldots,4$.
The representatives of the cohomology classes with $g=1,\ldots,4$
can be chosen such that they
do not depend on antifields at all. Due to their special
importance we summarize and comment only the results for $g=-1,0,1$
in detail.

The cohomology classes with $g=-1$ correspond
one-to-one to the independent solutions $f^\m(X)$ of \Gl{r9}
which can be interpreted as the Killing vectors in target
space. Each of them
generates a global symmetry of $S_0$ according to \Gl{r10}.
The resulting BRST--invariant functionals are given by \Gl{r11}.
This result is not surprising since it has been shown in \cite{bbh}
that the BRST cohomology classes with
ghost number $-1$ correspond one-to-one to the independent
nontrivial continuous global symmetries of the classical action
which is part of a cohomological
reformulation of Noether's theorem.

For $g=0$ there are two types of cohomology classes. Those of the
first type are represented by antifield--independent functionals
and provide the most general classical action for models
characterized by (i)--(iv). It is given by \Gl{r1}, with
the understanding that two such actions are equivalent if they
are related by a target space reparametrization \Gl{transfGB}.
Representatives of cohomology classes of the second type
depend nontrivially on the antifields. They
correspond one-to-one to those Killing vectors $f^\m(X)$
which satisfy \Gl{r12}. The corresponding BRST--invariant functionals
are given by \Gl{r12a} (and an analogous expression for $\cM_-$).
They correspond to so-called background charges and might provide
BRST--invariant functionals $\cM=\cM_++\cM_-$ which, as
remarked in the introduction, can
be used in order to look for a consistent extension of the
models or investigate an anomaly cancellation through background
charges (these applications would
require appropriate choices of $G_{\m\n}$, $B_{\m\n}$ and $f^\m$).

The cohomology classes with $g=1$
represent candidate anomalies. One can distinguish two types of them.
Representatives of the first type can be chosen to be
independent of the matter fields. In fact there are
precisely two inequivalent cohomology classes of this type,
represented by the matter field independent
functionals $\cH_+$ and $\cH_-$ given in eq. \Gl{r5} (contrary
to slightly misleading formulations in \cite{OSS} which
give the impression that there is only one cohomology class
represented by a special linear combination of $\cH_+$ and $\cH_-$).
Candidate anomalies of the second type depend nontrivially on the
matter fields and are represented by the functionals \Gl{r6}.
A functional \Gl{r6} is cohomologically trivial
if and only if $f^\pm_{\m\n}$ have the form \Gl{r8}.
All other functionals \Gl{r6} are BRST invariant
and cohomological nontrivial in the
complete space of local functionals of fields and antifields.
This result corrects a statement given in \cite{bala}
where the authors claim that matter field dependent
contributions to
BRST--invariant functionals with ghost number 1 can be
always removed by adding trivial contributions.
It is worth noting in this context that in fact both types of
anomalies arise in a generic model.
The requirement that the matter field dependent anomalies
vanish at the one-loop level
imposes the target space ``equations of motion'' for $G_{\m\n}$
and $B_{\m\n}$, vanishing of the matter field
independent anomalies fixes the target space dimension to $D=26$,
as discussed e.g.~in \cite{friedan}
(the quantities corresponding to
the symmetric and antisymmetric parts of
$f^\pm_{\m\n}$ and to $a_\pm$ occurring in the sum
${\cal H}_++{\cal H}_-$ for $a_+$=$a_-$
are in the second ref.~\cite{friedan} denoted by $\be^G_{\m\n}$,
$\be^B_{\m\n}$ and $\be^\PH$ respectively).

Finally we point out that
the absence of anomaly candidates depending nontrivially on the
antifields is a general feature of all models characterized by
(i)--(iv) and represents a remarkable difference to the situation
in Yang--Mills and Einstein--Yang--Mills theories with a
gauge group containing at least two
abelian factors if the classical action has at least one nontrivial
global symmetry \cite{anti}.

\end{document}